 \documentclass[preprint]{revtex4} 

 \usepackage{graphicx}

\usepackage{amssymb}

\begin{document}



\title{ Computation of Growth Rates and Threshold of the Electromagnetic Electron 
Temperature Gradient Modes in Tokamaks}
\author{Varun Tangri } 
\email{varun.tangri@gmail.com}



\begin{abstract}
In this manuscript, eigenvalues of the  
Electron Temperature Gradient (ETG) modes and Ion Temperature Gradient (ITG) modes are determined 
numerically using Hermite and Sinc differentiation matrices based 
 methods. It is shown that these methods are very useful for the computation of growth 
rates and threshold of the ETG  and ITG modes. The total number of accurately 
computed eigenvalues for the modes have also been computed.
 The ideas developed here are also of relevance to other modes that use Ballooning formalism.
\end{abstract}

\maketitle
\section{Introduction}
\label{intro}
Computation of growth rates of microinstabilities in tokamaks is of considerable interest to
the controlled fusion community. Often nonlocal analysis of Braginskii, Gyrokinetic or Gyrofluid
equations leads us to a system of differential equations or differentio-integral equations 
with an unknown 'eigenvalue'. Such eigenvalue problems are encountered extensively in areas of
 ballooning modes, drift waves and other microinstabilities.
  Such equations have usually been solved by 
the shooting method, integral
eigenvalue codes, phase integral methods and so on
 by a number of authors  and Refs \cite{Jhowry}-\cite{Raj} is certainly an incomplete list.

However shooting methods suffer from a number of disadvantages. 
 Shooting methods are very sensitive to initial 
conditions. It may be difficult to
arrive at the right answer unless a very good approximation to the correct result is known.
Furthermore, if determination the full eigenmode spectrum is intended then the whole 2-D parameter 
space of initial  guesses $\omega_r + i \omega_i$ must be spanned. 
Keeping in mind sensitivity to the initial guess, this may not be easy.
Shooting methods can run into difficulties if the eigenvalues of the
Jacobian matrix are widely separated.
Often results are determined by round off errors rather than by the equation itself.  
Integral eigenvalue codes like those that use methods of Delves and Lyness or Davies Method
also require initial guesses \cite{Antia}. 

Two methods that overcome many of these disadvantages are the Finite Difference (FD) and
Spectral collocation methods \cite{BoydBook}. Finite difference methods consist of replacing the 
derivatives
by an approximation ( such as $du/dx \approx [u(x+h) - u(x-h)]/2h$ where $h$ is the 
grid size. This introduces truncation error and accuracy may be compromised (although a finite difference method
is easy to program). Higher order finite difference
methods impose strong stability restriction. Gary and Hegalson \cite{gary} have highlighted the usefulness
of difference methods in computing eigenvalues of differential equations. Specifically, it has been
pointed out by them that the shooting methods that use Muller method or Lagurre method 
may run into difficulties (see table VI of Ref \cite{gary}).

Spectral methods involve expansion of the solution by a finite sum in a orthogonal
basis functions and minimizing of the residual \cite{BoydBook}. Different spectral 
and pseudospectral methods
differ in their minimization strategies. In this manuscript,  the use of
Differentiation Matrix (DM) based spectral methods to compute the ETG 
growth rate has been illustrated.
Differentiation matrix based methods score over 
finite difference methods in their ease of use, greater accuracy and faster
execution for the same matrix size. 
The importance of these methods was realized by the author
through an informal comparison observed in course of this work: a $15 
\times 15$ DM gave similar accuracy as a $150 \times 150$ FD matrix.
Since computational time scales very fast (It costs $\sim 10 N^3$ 
operations for the QZ \cite{Numrecipes, QZAlg} algorithm), 
the ratio of the time consumed in the FD
method ($N= 150$) while that in DM ($N=15$) would be $(150)^3/15^3=1000$!!
- a huge increase in speed!. It must be noted however, the matrices
involved in the DM and FD methods are, respectively, banded
and full. This might make the DM method much more expensive
for the same N number. Moreover, the FD performance could
probably also be improved by optimizing the non-equidistant grid.

Furthermore, as will be shown in this report, ETG growth rates converge faster 
 with differentiation matrix methods.
Differentiation matrix methods have one more advantage. The process of differentiation can be 
done through the matrix-vector product $f^{\ell}=D^{(\ell)} f$ where 
$f$ is a vector of function values at the nodes $\{x_k\}$. For example consider the model problem
\[  d^2 f/dx^2 - d f/dx = \lambda f  \]
 which in matrix form becomes  
\[
D^{(2)} f -D^{(1)} f = \lambda f,
\]
where $D^{(1)}$ and $D^{(2)}$ are a first and
second order differentiation matrices respectively. Thus in these methods, the process of differentiation can be 
carried out through a matrix-vector product. Thus in principle one could create differentiation matrices for 
the curl or a gradient operator to solve more complex problems.
The application of differentiation
matrices is relavant to plasma physics and specifically to toroidal
version of drift instabilities employing the ballooning formalism.

It must be pointed out that the goal of this manuscript is not to
show that finite difference algorithms are not useful. In fact, a variety 
of situations can be envisaged where finite difference algorithms may be indispensible.
Furthermore, neither the finite difference algorithm nor the differentiation
matrix based methods used here are the very
best and therefore a full academic comparison therefore cannot be done.
Since a general analytical solution of the equations discussed here in not available, 
a comparision of numerical solutions from FD and DM methods is used as a means to fill this gap. 

The main goal therefore, is to present application of differentiation
matrices to linear analysis of ballooning modes.
In the present paper, features of the collisionless, linear
electromagnetic ETG mode\cite{TangriETG, HortonETG}  are investigated used as a specific expample
of ballooning type modes. The ETG mode is widely recognized
as a mechanism for the small scale turbulence driven electron heat transport
in tokamaks. The ETG mode is a
short wavelength ($\rho_i>>\lambda_{\perp} >> \rho_e$), low frequency
($\omega << \Omega_e$) fast growing mode
($\gamma_{ETG}$ $\simeq \sqrt{\epsilon_{n} \eta_{e}} c_{e} /L_{n}$) driven 
unstable by temperature and density gradients. Here, $\rho_i$ is the ion gyroradius,
$\rho_e$ is the electron gyroradius, $\Omega_e$ is the electron cyclotron frequency,
$\epsilon_n=2L_n/R$, $\eta_e=L_n/L_{Te}$ where $L_n$ is the background density scale length.

\begin{figure}[p]
\begin{center}
\includegraphics[page=1, bb=96   189   549   611]{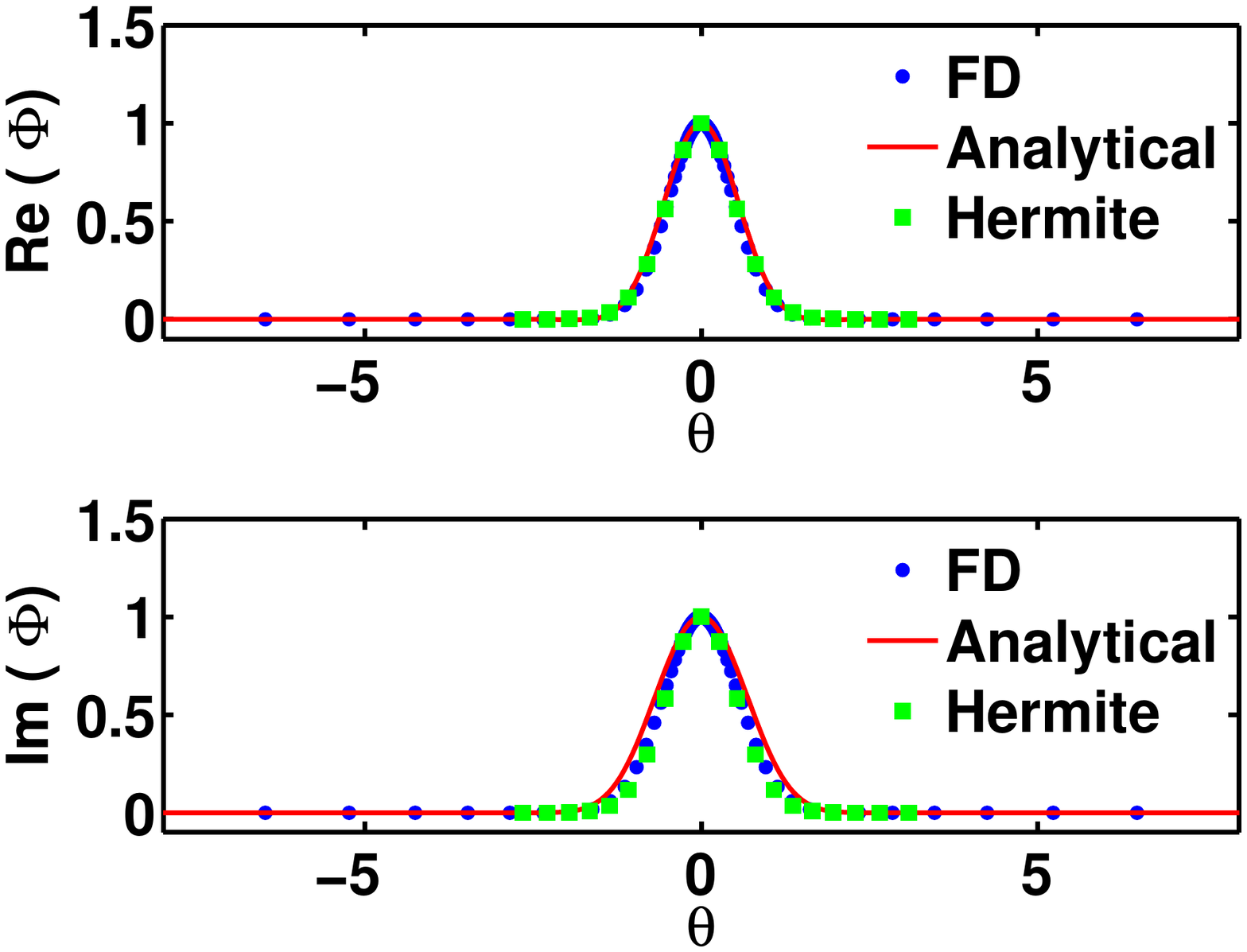}
\end{center}
\caption{ \label{eigvect}}
\end{figure}

The organization of this paper is as follows.
In the next section (Sec \ref{ETG}), the basic equations of the ETG mode will be presented. 
An approximate solution in the strong ballooning limit is also derived and approximate
eigenvectors computed (Figure \ref{eigvect}). Finite difference and spectral collocation methods
form the subject of the sections \ref{sFD} and \ref{sDM}.
Section \ref{sFD}  
presents the finite difference method used for benchmarking. An optimised finite difference method is also presented.
 Next, in Sec. \ref{sFD}, comparisons
with the 'analytical approximation' (Figure \ref{eigvect}) is also presented. Disadvantages of the 
method are also highlighted. In section \ref{sDM},
the differentiation matrix based method are compared with the finite difference method.
For this purpose, comparison is made in three different ways: (i) Growth, (ii) Matrix size, and (iii)
Conjugacy of eigenvalues. Next, the problem of determination of the number of eigenvalues is addressed
for which the 'nearest' and 'ordinal' distances have been computed.  
A summary and conclusions are presented in Section \ref{Concl}.

\section{The basic equations of electron temperature gradient mode \label{ETG} }
In this section the fluid description of the ETG mode will be discussed and a set of basic 
equations will be presented. An advanced electron fluid model for the core
plasma (i.e. collisionless plasma) has been used. This fluid model is analogous to 
the fluid equations used in Refs. (\cite{SinghNF}, \cite{TangriETG}, \cite{mypop}).
 The governing model equations are electron continuity 
equation, parallel momentum-equation and temperature
equation:  
\begin{equation} \label{FLeq}
\begin{array}{l}
 \omega \left[ {\tau  - \nabla _ \bot ^2 } \right]\tilde \phi  + \left[ {1 - \tilde \varepsilon _n (1 + \tau ) + (1 + \eta _e )\nabla _ \bot ^2 } \right]k_\theta  \tilde \phi  \\
+  \tilde \varepsilon _n k_\theta  \tilde T + i\frac{{L_n }}{{qR}}\frac{\partial }{{\partial \theta }}(\tilde A_{||} k_ \bot  ) = 0, \\
 ( - (\frac{\beta }{2} - \nabla _ \bot ^2 )i\omega  + \frac{\beta }{2}(1 + \eta
_e )ik_\theta  )\tilde A_{||}  + \frac{{\varepsilon _n }}{{2q}}\frac{\partial }{{\partial \theta }}(\tilde \phi  - \tilde n - \tilde T) = 0, \\
  - \omega \tilde T + k_\theta  \frac{5}{3}\tilde \varepsilon _n \tilde T + \left[ {(\eta _e  - 2/3)k_\theta   - \frac{{2\tau }}{3}\omega } \right]\tilde \phi
= 0. \\
 \end{array}
\end{equation}
Where $\tilde n$, $\tilde \phi, \tilde T$ and $\tilde A_{||}$ are the normalized perturbations in 
density $\delta n$, potential ($\delta \phi$) , electron temperature
($\delta T_e$) and parallel vector potential ($e A_{||}/T_e$) respectively: $\left(\tilde{\phi},\tilde{n},\tilde{T},\delta\tilde{B}_\parallel \right) =
L_n/\rho_e \left(e \delta \phi/T_{e0},\delta n/n_0,\delta T_e/T_{e0},
\delta B_\parallel/ B \right)$, $\tilde{A}_\parallel = (2 c_e L_n/\beta_e
c \rho_e)e A_\parallel/T_{e0}$. The perpendicular scales are normalized to $\rho_e$, 
the electron larmor radius while the parallel scale is normalized to $L_n$. As in the standard 
ballooning formalism, 
$\nabla_\perp^2 f = - k_\perp^2 f = - k_\theta^2 f \left(1 + (s \hat
{\theta} - \alpha \sin \hat{\theta})^2 \right), 
\tilde{\epsilon}_n = \epsilon_n \left[\cos \theta + \left(s \theta - \alpha
\sin \theta \right) \sin \theta \right]$, $
\nabla_\parallel f = i k_\parallel f \simeq \left(1/qR\right)\left(\partial f/
\partial \theta \right)$
where $\theta$ is the extended coordinate in the ballooning formalism. Here $\omega$ is the 
unknown, complex eigenvalue to be determined and $\epsilon_n=2 L_n/R$,
where $R$ is the major radius and $L_n$ is the density scale length. Also note that this is a set of coupled, complex,
differential and algebraic equations.

First, an approximate analytical solution for a qualitative comparison is presented.
A general analytical solution of Eq. \ref{FLeq} is not available, however the most 
important case where one can obtain an analytical solution is the
strong ballooning limit. In the strong ballooning limit, one can assume that
$g(\theta)=cos(\theta)+(s \theta-\alpha sin(\theta))^2$ is slowly varying with $\theta$ 
so that $g(\theta)
\simeq g(0) = 1$. In this limit, Eq. \ref{FLeq} can be written as
\begin{equation} \label{HermEQ}
A \frac{\partial^2 \tilde{\Phi}}{\partial \theta^2} = (B + C \theta^2) \tilde{\Phi}
\end{equation}
%
\begin{equation} \label{defA}
A = \frac{\epsilon^2_n}{4 q^2 \hat{\omega}} \left[ \left( \eta_e - \frac{2}{3}
\right) k + \frac{5}{3} (1 + \tau^*_0) \epsilon_n\ k - \left(1 + \frac{5}{3}
\tau_0^* \right) \hat{\omega} \right]
\end{equation}
\begin{eqnarray}  \label{defB}
B = \hat{\omega}^2 \left( \tau_0^* + k^2 \right) + \hat{\omega} \ k \left[ 1
- \epsilon_n \left( 1 + \frac{10}{3} \tau^*_0 \right) - k^2 \left( 1 + \eta_e
+ \frac{5}{3} \epsilon_n \right) \right] \nonumber\\[0.1in]
+ \epsilon_n k^2 \left[ \eta_e - \frac{7}{3} + \frac{5}{3} \epsilon_n (1 +
\tau^*_0) + \frac{5}{3} k^2 (1 + \eta_e) \right]
\end{eqnarray}
\begin{equation}  \label{defC}
C = k^2 \hat{s}^2 \left[ \hat{\omega}^2 - \hat{\omega} k \left( 1 + \eta_e +
\frac{5}{3} \epsilon_n \right) + \frac{5}{3} \epsilon_n k^2 (1 + \eta_e)
\right]
\end{equation}
%
where $\tau^*_0 = k^2_{\theta} \lambda^2_{De} + \tau + \delta, \, \hat{\omega} = \omega L_n/c_e$ and $k = k_{\theta} \rho_e$.
One  specially looks the solutions of  Eq. \ref{HermEQ} of the 
form 
\begin{equation} \label{AnalyticalVector}
\Phi(\theta) =  \exp (- \sigma \theta^2/2), 
\end{equation}
where $Re(\sigma) > 0$, where $\sigma = \pm \sqrt{C/A}$, this gives us the dispersion relation
\begin{equation}  \label{defDISP}
 B = - \sigma A.
 \end{equation}
Here $A,B$ and $C$ are parameters independent of $\theta$ as in Ref \cite{mypop}. Note that the solutions of the form
$\Phi(\theta) \propto F(\theta) exp(-\sigma \theta^2)$ were used. Especially the case $F(\theta)=1$, which corresponds 
to the above solution.

 Next, a discussion of the numerical solution of the equations [\ref{defA}-\ref{defDISP}] will be discussed. 
This will
be used as a verification benchmark for the finite difference and differentiation matrix based methods
of solution of Eqs. (\ref{FLeq}). The function $h(\hat{\omega})=B+ \sigma A$ is a nonlinear function of the 
complex variable $\hat{\omega}=\omega_r + i \omega_i$, hence we have solved Eq. (\ref{defDISP})
 by finding the roots of the  function $h(\hat{\omega})$ using Muller method.
In figure \ref{eigvect}, the eigenvectors corresponding to $F(\theta)=1$ from the above
'analytical' solution have been drawn as solid lines.
The dots in this figure will form the subject of our next discussion - the finite difference method, which is discussed Sec. \ref{sFD}.
The solid squares in this figure refer to computation using the Hermite Method as descibed in Sec. \ref{sDM}.
Please note that in this figure, for easy comparison the eigenvectors have been normalized in such 
way so that $Real[\Phi(\theta=0)]=1$ 
and $Imag[\Phi(\theta =0)]=1$ for all the methods.
%
\section{Finite Difference  Methods\label{sFD} }
The basic equations solved in this manuscript, Eq. (\ref{FLeq}), can be considered as
 a system of first order ordinary differential equations of the form 
\begin{equation}   \label{geneq} 
\frac{d y_i}{dx}+ \sum b_j(x) y_j=\lambda \sum a_j(x) y_j
\end{equation}
Often, in the past, a number of authors have solved similar eigenvalue problems but as a single 
second order differential equation that involves powers of the eigenvalue $\lambda$ in the following manner:
\begin{equation}   \label{geneq} 
a(x,\lambda) \frac{d^2 y}{dx^2}= \sum b(x,\lambda) y
\end{equation}
where $a(x,\lambda)$, $b(x,\lambda)$ can be nonlinear functions in $x$ but polynomial
in the eigenvalue $\lambda$. 

Finite difference methods are the most widely used methods to solve for
eigenvalues of such ODE-eigenvalue problems. The basic ideology that we have used is highlighted in the following
equations:
\begin{eqnarray} \label{FDeqn}
\frac{y_k-y_{k-1}}{x_k-x_{k-1}}+ \sum b_j(\frac{1}{2}(x_k+x_{k-1})) (y_k+y_{k-1})/2  \\
=\lambda \sum a_j(\frac{1}{2}(x_k+x_{k-1})) (y_k+y_{k-1})/2  \nonumber
\end{eqnarray}
An implementation of such finite difference methods is explained in Ref \cite{Antia}.  Although the
 method is elaborate, it has the following  
disadvantages. This method computes the eigenvalues by 
calculating the zeros of  a determinant.
These zeros are computed using an iterative process which in turn, requires guesses.
Requirement of  guess values that makes it no better than shooting methods discussed earlier.
Furthermore, for efficient use, the basins of attraction of the various initial guesses
must be known beforehand.

Instead of using the above code, a finite difference code 
has been developed to overcome the above shortcomings. This 
code has the following advantages not present in the above code:
(i) generates the full eigenvalue spectrum simultaneously;
(ii) there is an option to refine a particular eigenvalue;
 and (ii) for small changes in a parameter, the changes in a branch
can be tracked automatically, otherwise it requires careful human intervention.
Other salient feature are: (i) Can solve complex coupled differentio-algebraic equations; and (ii) arbitrary mesh (which is necessary and
useful since boundary conditions are at infinity and the eigenfunction is expected to be 
localized).

Next, the finite difference numerical scheme used is explained in detail. First, an odd number of
grid points such that $\theta=0$ is the central point $I_o$ is chosen. Next, allocate grid points
such that $\theta (I_o+j)=-\theta (I_o-j)$. The points $\theta (I)$ need not be equally spaced
but can be decided according to the problem at hand. The allocation of grid points was not 
optimized but chosen intuitively such that they are either (i) equally spaced; or (ii)
 $ \theta \propto F(\theta_1)$ where $\theta_1=0,\delta \theta, 2 \; \delta \theta,......,0.99$
is a set of equally  spaced points between 0 and 0.99.  
The function $F(\theta_1)$ can be any monotonically increasing function of $\theta_1$
like $1/(1-\theta_1)$, $exp(c \theta_1)$ and so on. Next one discretizes the two differential 
equations at half grid points and the algebraic equation at full grid points. All eigenvalues and eigenvectors can be 
obtained simultaneously if the resulting 
set of coupled algebraic equations are solved by the QZ algorithm.

 In the strong ballooning limit, the full numerical computation
closely follows the analytical results. In Figure \ref{eigvect}, eigenvectors
from both the analytical and numerical computations are depicted as a validation of the code.
The solid lines indicate analytical result while the solid circles depict the numerical values. It must
also be pointed out that in this figure, the x-axis range has been reduced (from the actual range used in the computation) to show that the eigenvectors 
from the analytical
approximation closely follow the numerical result. The non-uniform distribution of grid points
 just explained is also illustrated in Figure \ref{eigvect} as solid dots.

This finite difference method is not without disadvantages. Eq. (\ref{FLeq})
has three equations (two ODE's and one algebraic equation). Thus even for a modest 
mesh size of 150 points generates two matrices each of size $(3*150) \times (3*150)$.
Investigation of the ETG mode requires variation of the basic parameters over a wide range.
 For example, for each value of shear parameter 
$\hat{s}$ and ballooning parameter  $\alpha$, safety factor $q$, plasma beta $\beta$, temperature ratio $\tau$,  eigenvalues for at
 least $200-300$ values of $\eta_i$ must be computed. This large number of values of $\eta_i$ are used to
accurately compute the threshold and this increases the time taken by the computer.

\section{Eigenvalues using Differentiation matrices \label{sDM} }
Differentiation matrices are novel, powerful methods of finding the eigenvalues of
differential equations. These methods are very 
accurate. For smooth problems, convergence rates of the order of $O(e^{-cN})$ are often 
achieved \cite{Canuto, Tadmor} as compared 
to finite difference methods where the convergence rate are slower, 
typically ($N^{-2}$) or ($N^{-4}$). 

The method employed is briefly explained below:
A spectral collocation method \cite{Weid} for solving differential equations that is based on 
weighted interpolants of the form:
\begin{equation}
f(x) \approx \sum_{j=1}^N \frac{\alpha (x)}{\alpha (x_j)} \phi_j(x) f_j,
\end{equation}
has been used.
The points $\{x_j\}_{j=1}^N$ is a set of distinct interpolation nodes, $\alpha(x)$ is a weight 
function, $f_j=f(x_j)$ and the set of interpolating functions $\{ \phi_j(x)\}_{j=1}^N$ satisfies
$\phi_j(x_k)=\delta_{jk}$. Where $\delta_{jk}$ is the Kronecker delta function. The differentiation 
 matrix is then defined as a matrix with the entries:
\begin{equation}
D_{k,j}^{(\ell)}=\frac{d^\ell}{dx^{\ell}} [ \frac{\alpha (x)}{\alpha (x_j)} \phi_j(x)  ]_{x=x_k}
\end{equation}
The process of differentiation can be done through the matrix-vector product $f^{\ell}=D^{(\ell)} f$ where 
$f$ is a vector of function values at the nodes $\{x_k\}$. 

This manuscript, mostly uses the Hermite Collocation method.
In the Hermite Collocation method, the nodes $x_1, x_2, x_3,....x_N$ are the roots of $H_N (x)$, the Hermite polynomial.
The weight function is the Gaussian function $\alpha (x)=exp(-x^2/2)$ and the interpolant is defined as
\begin{equation}
p_{N-1}(x)=\sum_{j=1}^N\frac{exp(-x^2/2)}{exp(-x_j^2/2)} \phi_j(x) f_j,
\end{equation}
where,
\begin{equation}
\phi_j(x)=\frac{H_N(x)}{H'_N(x_j)(x-x_j)}.
\end{equation}
It must be noted that the real line $(-\infty, \infty)$ may be mapped to itself by a change of variable 
$x=b \tilde{x}$, where $b$ is any positive real number. Thus the freedom offered by the parameter $b$ 
may be exploited to optimise the accuracy of the Hermite differencing process.
Before proceeding further let us discuss the solution of Hermite
differential equation itself. The solution the Hermite differential equation using a Hermite DM was performed. 
The errors in the first twenty five eigenvalues was found to be less than $10^{-8}$.

The application of this method to plasma physics is not new. It has been used in simulations of Vlasov 
equation since the sixties. Low order Hermite series ($N=2,3$) were shown to suppress numerical instabilities
in a Vlasov simulation
 by Engelmann {\it et al.} \cite{Engel}. More  recently Shumer {\it et al.} have used velocity-scaled Hermite 
representations \cite{Schumer}. In this manuscript, we employ Hermite differentiation matrices
to compute eigenvalues of the ETG instability.

\subsection{ Comparison with FD: Growth Rate}
\begin{figure}[p] 
\begin{center}
\includegraphics[page=1, bb=56   188   546   592 ]{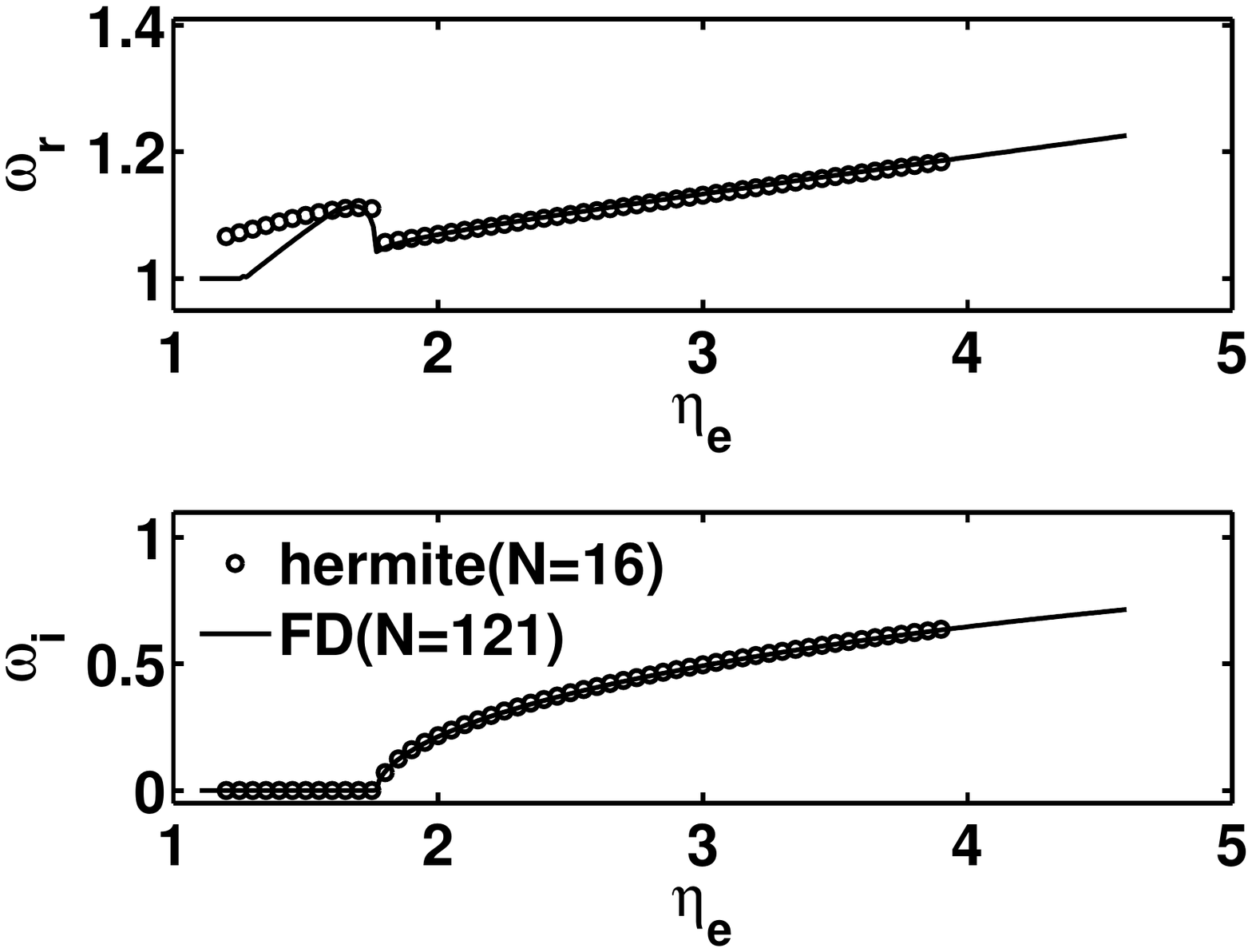}
\end{center} 
\caption{ \label{HN} } 
\end{figure}

The eigenvalue of the ETG mode has also been computed using the following basic parameters
$s=0.8, \tau=1, k_{\theta}=0.6,q=1.4, \epsilon_n=0.509$ and $\beta=0.01$, while $\eta_i$ is varied.
A typical result from these simulations is depicted in Figure \ref{HN}. In this figure, the results of 
computation from Finite difference method is shown as solid lines and that from Hermite method
is shown as circles. As can be seen in the figure, the Hermite method requires 
only $N=16$ to achieve accuracy comparable with a Finite difference computation 
with $N=121$. The finite difference code uses
an odd number of points $51, 121$ {\it etc.} so as to have equal number of points to the left and to the right
of $\theta=0$. Since the Hermite method requires only a small number of points, its is faster 
by orders of 
magnitudes. This allows greater parameter scans.


\subsection{ Comparison with FD: Matrix Size}
\begin{figure}[p]
\begin{center}
\includegraphics[page=1, bb= 75   201   552   589]{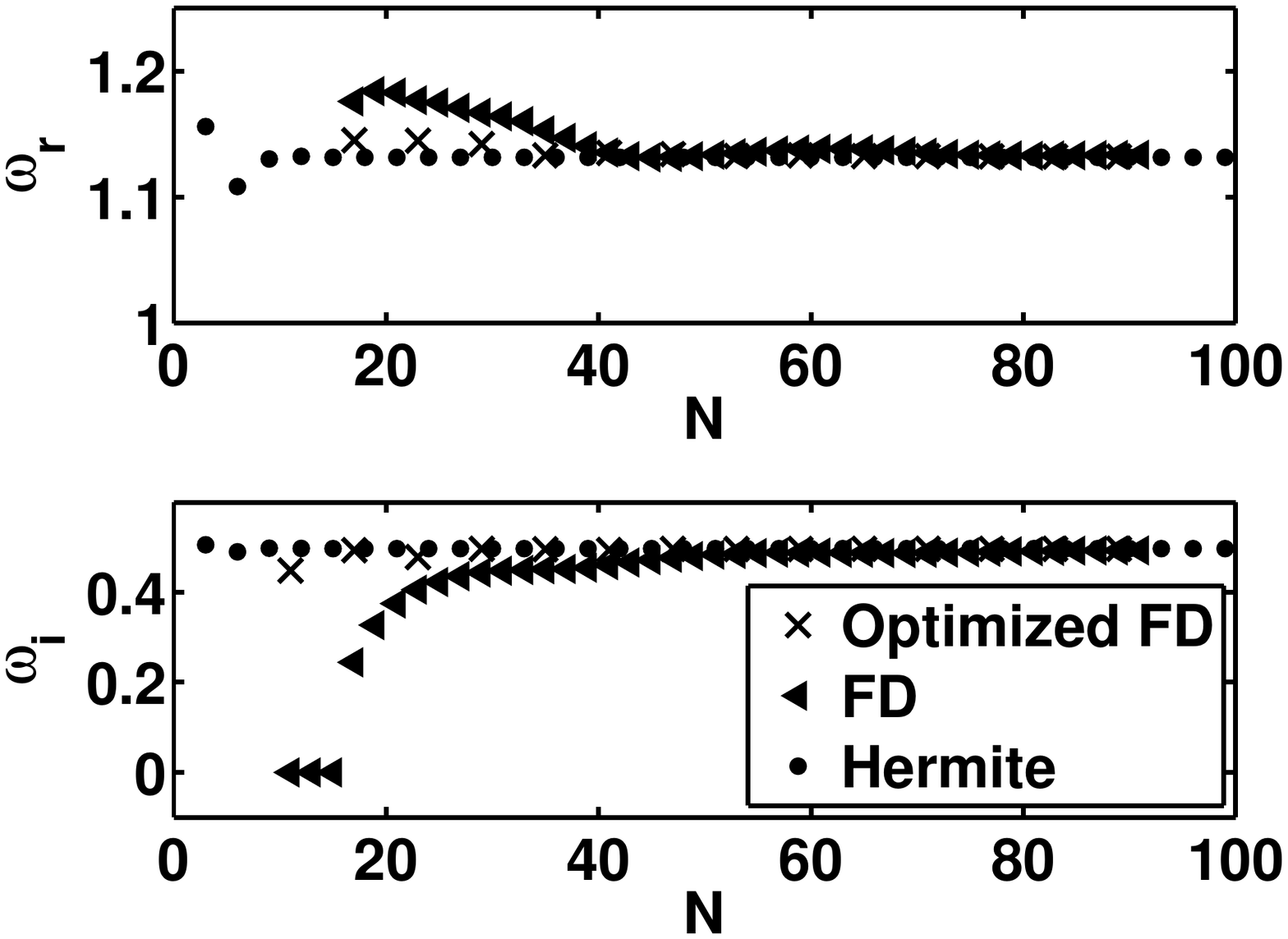}
\end{center} 
\caption{ \label{comparison}}
\end{figure}

The eigenvalues of the ETG mode for different values on N using the
finite difference method and Hermite spectral collocation method have also been computed. As 
depicted in Figure ~(\ref{comparison}), the Hermite method converges much before the two finite difference
methods. 
In Figure ~(\ref{comparison}), two finite difference methods are illustated by solid triangles and crosses.
These FD methods only differ in terms of how the mesh was optimized.
Below $N=20$, the finite difference code (which is illustated by triangles) gives wrong results
 ({\it i.e.,} $\omega_i \sim 0$), slowly this FD result begins to converge  and a 
stable value is obtained only beyond $N=100$. To improve the convergence of the FD method,
the mesh was optimized so that accuracy improves for $N <20$. This is shown as crosses in Figure~(\ref{comparison}).
The Hermite method, however gives us
an acceptable eigenvalue  even for $N=10$, or $N=15$. Note that the computational time
taken for an accurate computation would also be small since $N$ is small. 

\begin{figure}[p]
\begin{center}
\includegraphics[page=1, bb= 75   201   552   589]{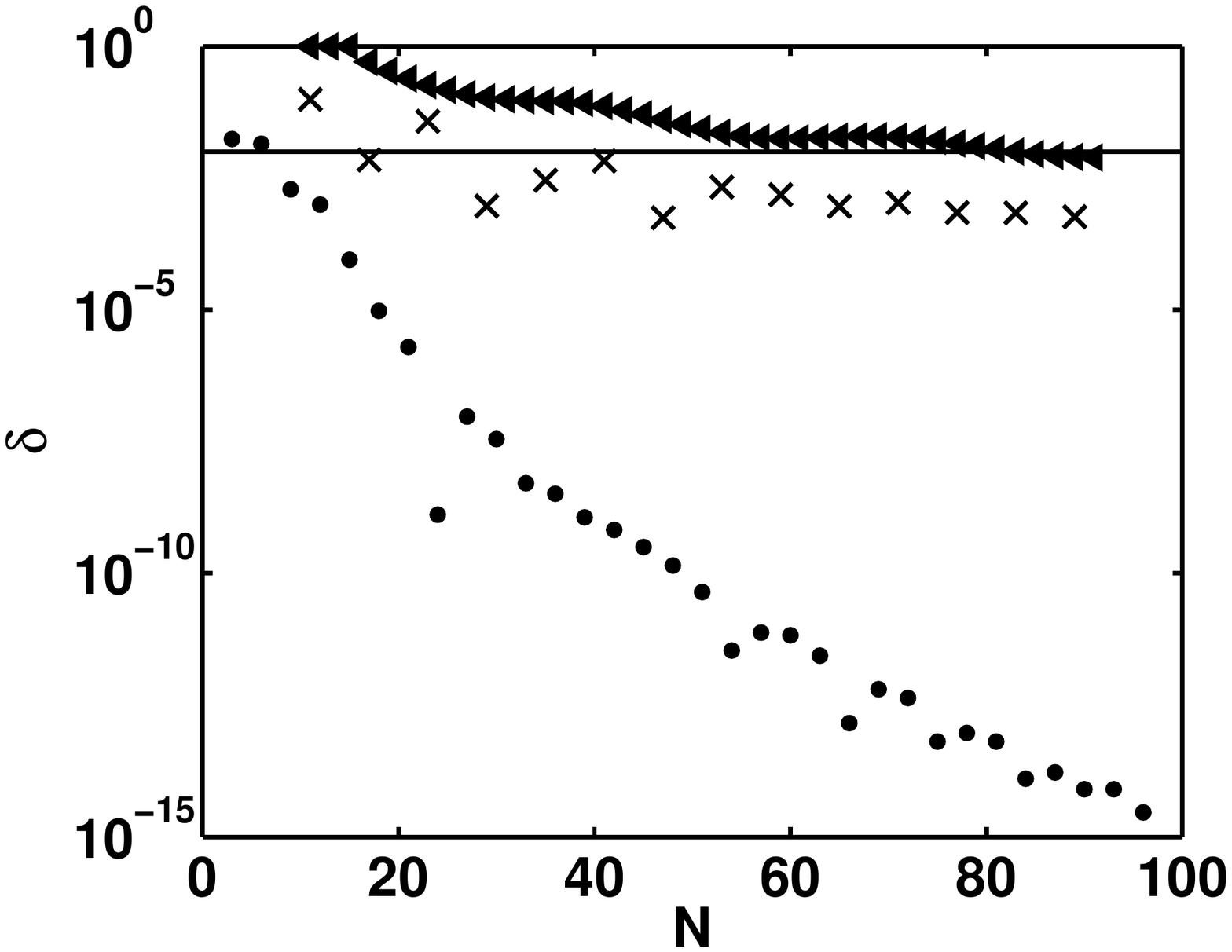}
\end{center} 
\caption{ \label{comparison2}}
\end{figure}

The rate of convergence of these three methods is illustated in Figure ~(\ref{comparison2})
where the error $\delta(N)$ is plotted versus $N$. Here,
\begin{equation}
\delta(N)= \frac{\omega_i(N) - \omega_i^h (N=100)}{\omega_i^h (N=100)}
\end{equation}
 and $\omega_i^h (N=100)$ is the growth rate from the Hermite method at $N=100$. 
In this figure, $\delta(N)$ from the unoptimised finite difference method is represented using solid triangles,
$\delta(N)$ from the  optimised finite difference method is represented by crosses and the $\delta(N)$ from 
the differentiation matrix (Hermite) method
is represented by solid dots. The solid line in this figure represents an error level of $1 \%$ which is easily
achieved by the Hermite method. The error also decreases fastest for the this method.
 

\subsection{ Comparison with FD: Conjugacy of Growth Rates}
\begin{figure}[p]
\begin{center}
\includegraphics[page=1, bb= 52   189   555   597 ]{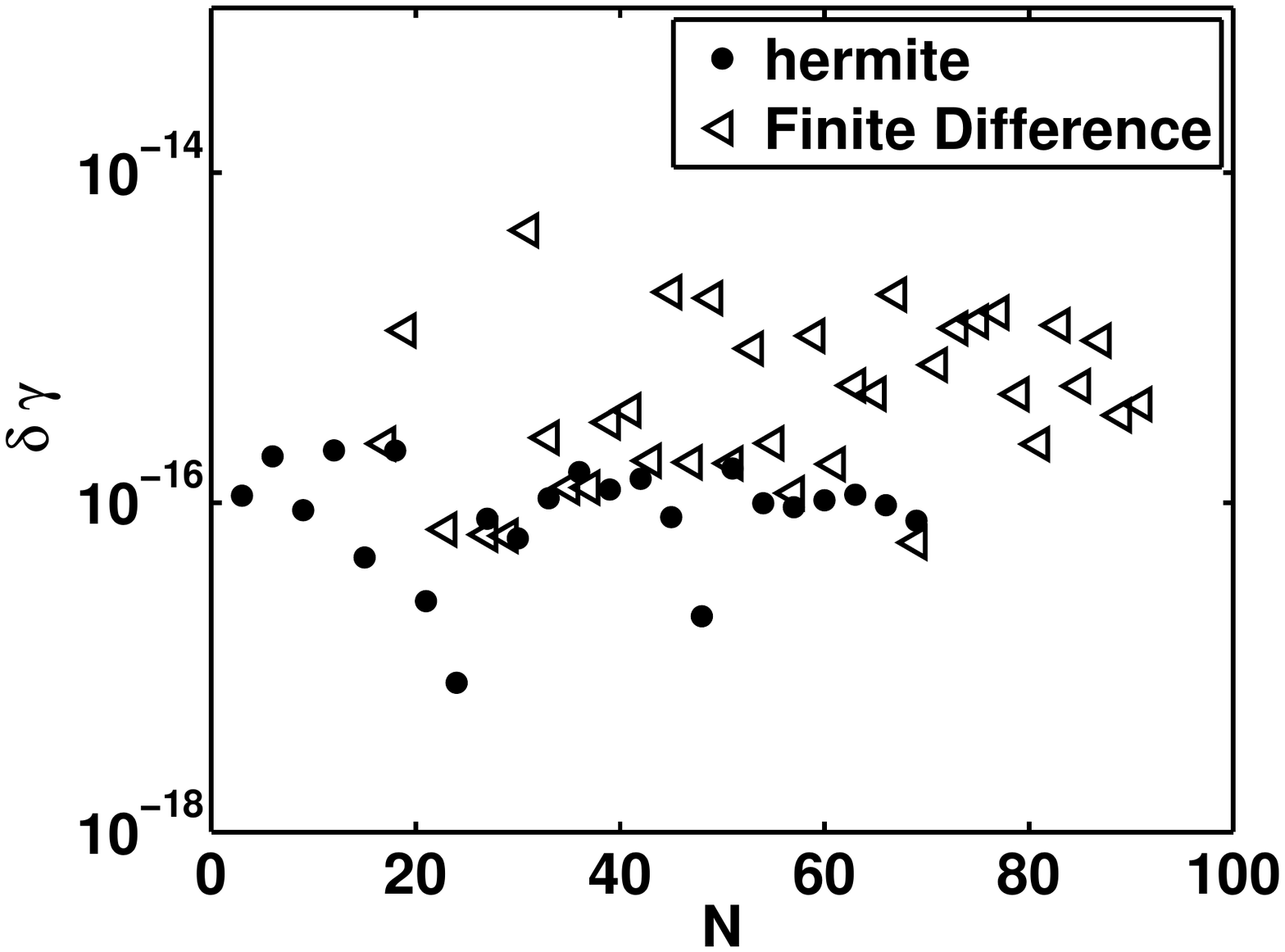}
\end{center} 
\caption{ \label{deltaGamma}}
\end{figure}

Looking at the basic equations, one realizes that for a particular set of parameters,
the maximum growth rate ($\gamma_{max}$) also has a complex  conjugate $\gamma_{min}$.
Mathematically, one expects that $\gamma_{max}=-\gamma_{min}$. This however may not
be true in reality. Only one of the two roots may be  more accurate than the other. Thus the difference
between the two growth rates can be considered to be a measure of the numerical errors in
computation. The parameter $\delta \gamma$ where, 
\begin{equation}
\delta \gamma = \frac{\gamma_{max}+\gamma_{min}}{\gamma_{max}-\gamma_{min}},
\end{equation}
can be considered as a measure of this error.
 To illustrate the
competitiveness of the method, the variation of $\delta \gamma$ for
various $N$, using the finite difference as well as the Hermite Method has also been computed. This 
is shown in 
Figure \ref{deltaGamma} where the triangles refer to a Hermite calculation while the 
dots refer to finite difference method. Values from both methods are small but the Hermite 
method appears to be slightly more accurate than finite difference. It is possible that the slightly 
smaller inaccuracies of the Hermite-case are simply due to
a slightly unfavorable ordering in the matrix or round off errors. It must also be noted that this error is also smaller 
compared to the error in the approximation
schemes.

\subsection{ Number of Convergent Eigenvalues}
The system of equations \ref{FLeq} has three equations in three variables. Increasing 
N, furthermore, would give us $3 N$ eigenvalues. For an arbitrary N, increasing to infinity, the
number of eigenvalues would also be large. All of these eigenvalues might not be good
 approximations to the eigenvalues of the actual differential equation.
 Physically one expects only a few of these should remain the same for every N.
Thus it is important to ascertain the number of true eigenvalues of the differential equation at hand.

  The following two situations can be envisaged:
(i) the eigenvalue 'drifts' as N increases. (ii) a new (probably spurious) eigenvalue is
 inserted between the two earlier ones. In order to ascertain the correct eigenvalues, Boyd {\it et. al.}
have proposed a scheme \cite{Boyd} for ocean tides bounded by meridians. 
Following this prescription, the 'ordinal' and 'nearest neighbor' drifts can be computed as explained below.

One computes eigenvalues using two different matrix sizes $N_1$ and $N_2$ such that $N_1/N_2 > 1/2$.
The physically important eigenvalues include the highest  growing ($\gamma_{max} >0$)
and the most stable  ($\gamma_{min} <0$)
These modes are often complex conjugates of each other ($\gamma_{max} \sim -\gamma_{min}$)  . 
Thus growth rate spectrum is then sorted in two different
 ways- ascending and descending. 

The ordinal drift is defined by 
\begin{equation}
\delta_{j,ordinal}= | \lambda_j^{N_1}- \lambda_j^{N_2} |,
\end{equation}
where $\lambda_j^{N}$  is the imaginary part of the  jth eigenvalue 
(after the eigenvalues have been sorted) as computed using N spectral coefficients.

While the 
nearest neighbor drift is given by
\begin{equation}
\delta_{j,nearest}=min_{k \epsilon [1,N_2]} | \lambda_j^{N_1}- \lambda_k^{N_2} |.
\end{equation}
next, the intermodal drift is defined as follows

\begin{eqnarray} \nonumber
\begin{array}{clcr}
\sigma_j=\left\{
\begin{array}{clcr}
 | \lambda_j^{N_1}- \lambda_j^{N_2} |,  &  if \; j=1 \\ { }\\
\frac{1}{2} \{ | \lambda_j- \lambda_{j-1} |+  | \lambda_{j+1}- \lambda_{j} | &  if \; j > 1,
\end{array}
\right.
\end{array}
\end{eqnarray}
Boyd {\it et. al.} then recommend making a log/linear plot of the following ratios:
\begin{equation}
r_{j,ordinal}=\frac{min ( | \lambda_j | ,\sigma_j )}{\delta_{j,ordinal}}
\end{equation}
\begin{equation}
r_{j,nearest}=\frac{min ( | \lambda_j | ,\sigma_j )}{\delta_{j,nearest}}
\end{equation}
which must both be large for the eigenvalue computation to be accurate. Boyd {\it et. al.}, illustrate
for Legendre's differential equation in Ref \cite{Boyd}, Figure 4.

\begin{figure}[p]
\begin{center}
\includegraphics[page=1, bb=74   189   551   597]{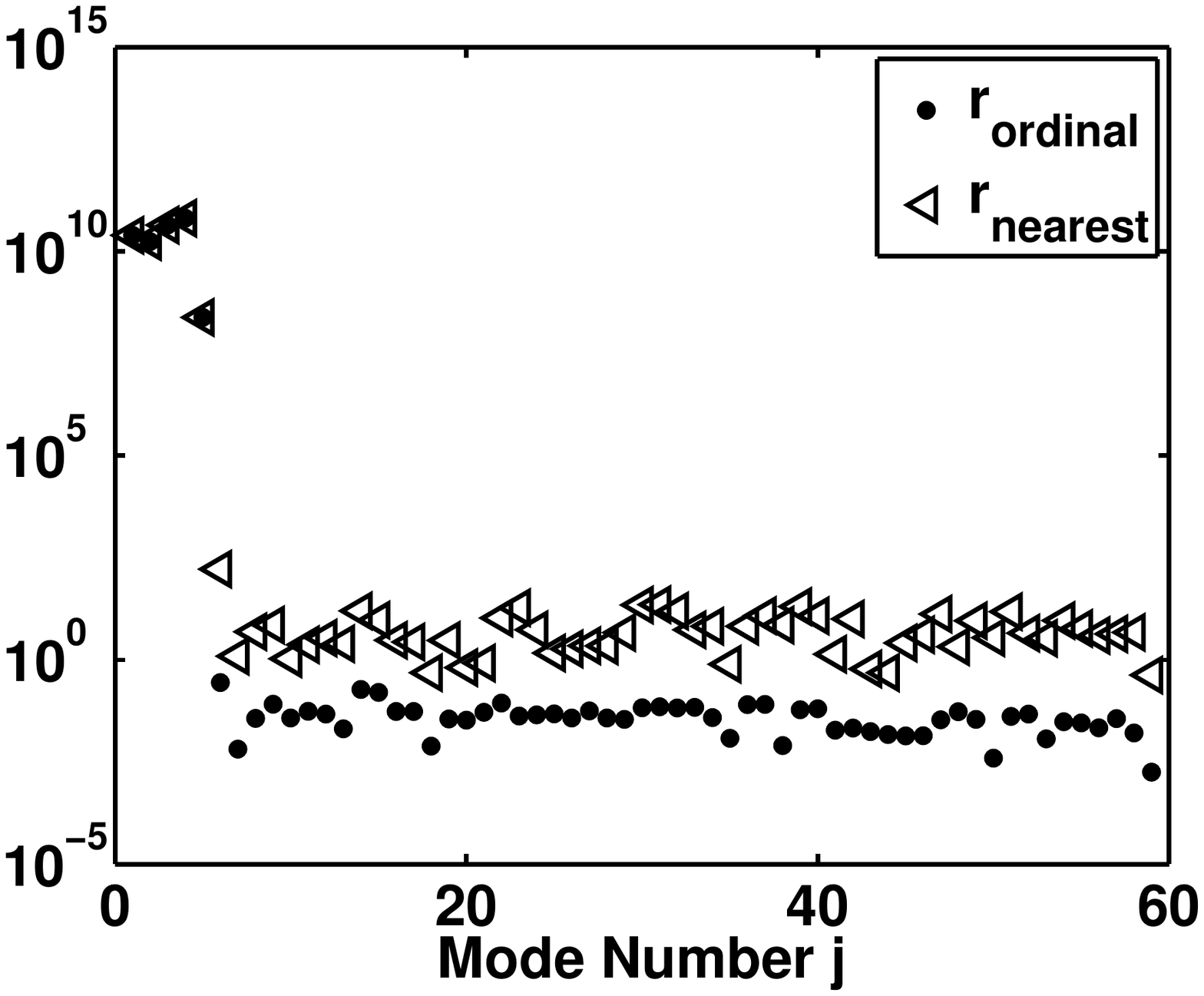}
\end{center} 
\caption{\label{en0p9} }
\end{figure}

The ordinal and nearest distances for the ETG mode has been computed using the 
parameters $s=0.8, \tau=1, k_{\theta}=0.6, q=1.4, \epsilon_n=0.909, \beta=0.01$ in Equation [\ref{FLeq}].
The choice of parameters is not arbitrary, they have been chosen to be the same as
those of Jenko {\it et. al.} \cite{Jenko}.
Next, the eigenvalues are  sorted in two different ways- ascending and descending orders. 
A total of ten numerically accurate eigenvalues were found, five each for each case
signifying five growing modes and five damped modes. 
The case of highest growing modes is illustrated in figure[\ref{en0p9}]. Such a large 
number of eigenvalues can be understood using the following argument.
The argument is based on the results of the local  and nonlocal analysis of the electron temperature
 gradient modes given in Ref. \cite{mypop} by R. Singh {\it et. al.} The Equation 31 of Ref \cite{mypop} 
is a third order dispersion relation retaining parallel dynamics. That is one growing, one stable and
and real frequency branch. Moving to nonlocal, Equations 41-44, we realize that the dispersion  
relation is now a fifth order. Furthermore Eqns 41-44 assume solutions of the type 
$exp(-\sigma \theta^2/2)$ and ignores the higher branches with the form
$H_n(\theta) exp(-\sigma \theta^2/2), \; n=1,2,3.....$ where $H_n$ is the Hermite polynomial.
Higher values of n add more branches to be found numerically. Thus the possibility 
 a large number of eigenvalues cannot be ruled out. This conclusion is not new. 
Gao {\it et. al.} \cite{gao} have identified a series of unstable branches for the ETG mode. 
Similarly \cite{dong} {\it et. al.} have shown six growing modes. Lee {\it et. al} [\cite{Lee} ]have 
studied the three most unstable modes as a function of parameters. 

\begin{figure}[p] 
\begin{center}
\includegraphics[page=1, bb=74   189   551   597]{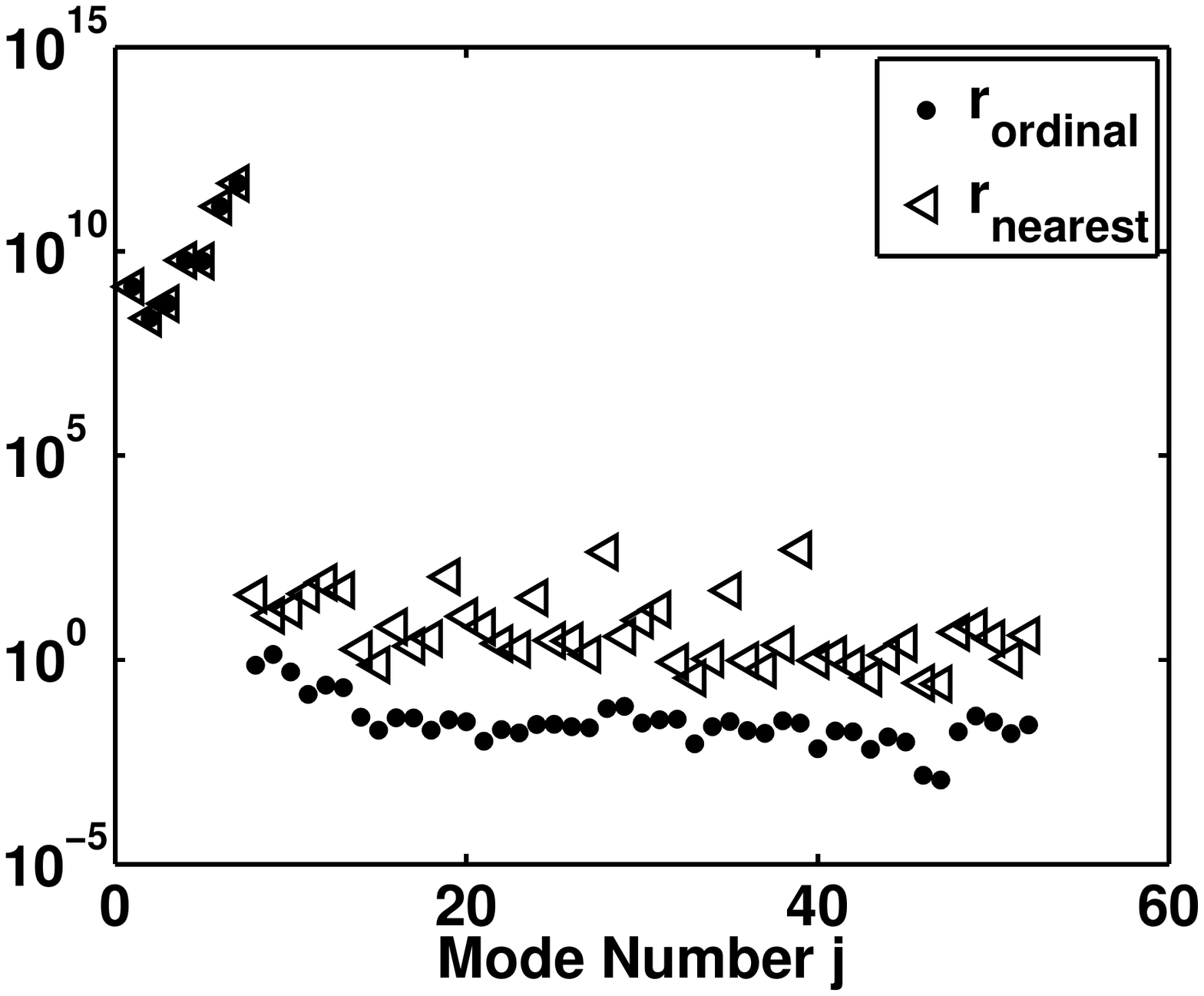}
\end{center} 
\caption{ \label{en0p5} }
\end{figure}

One might also ask oneself if the number of these eigenvalues changes with physical parameters. 
The variation of the number of good, highest growing eigenvalues with 
the parameter $\epsilon_n$ has aslo been investigated. For large values of  
$\epsilon_n$ ($\sim 0.9$), the number of good eigenvalues is five. However on 
decreasing $\epsilon_n$, ($\sim 0.5$), the number of good eigenvalues
increases and becomes seven.This is illustrated in figure \ref{en0p5}. 
 On a further decrease of  $\epsilon_n$, ($\sim 0.2$), the large separation 
between good eigenvalues and spurious diminishes with the mode number $j$.


\subsection{ The Sinc Method}
\begin{figure}[p] 
\begin{center}
\includegraphics[page=1, bb=55   188   546   592]{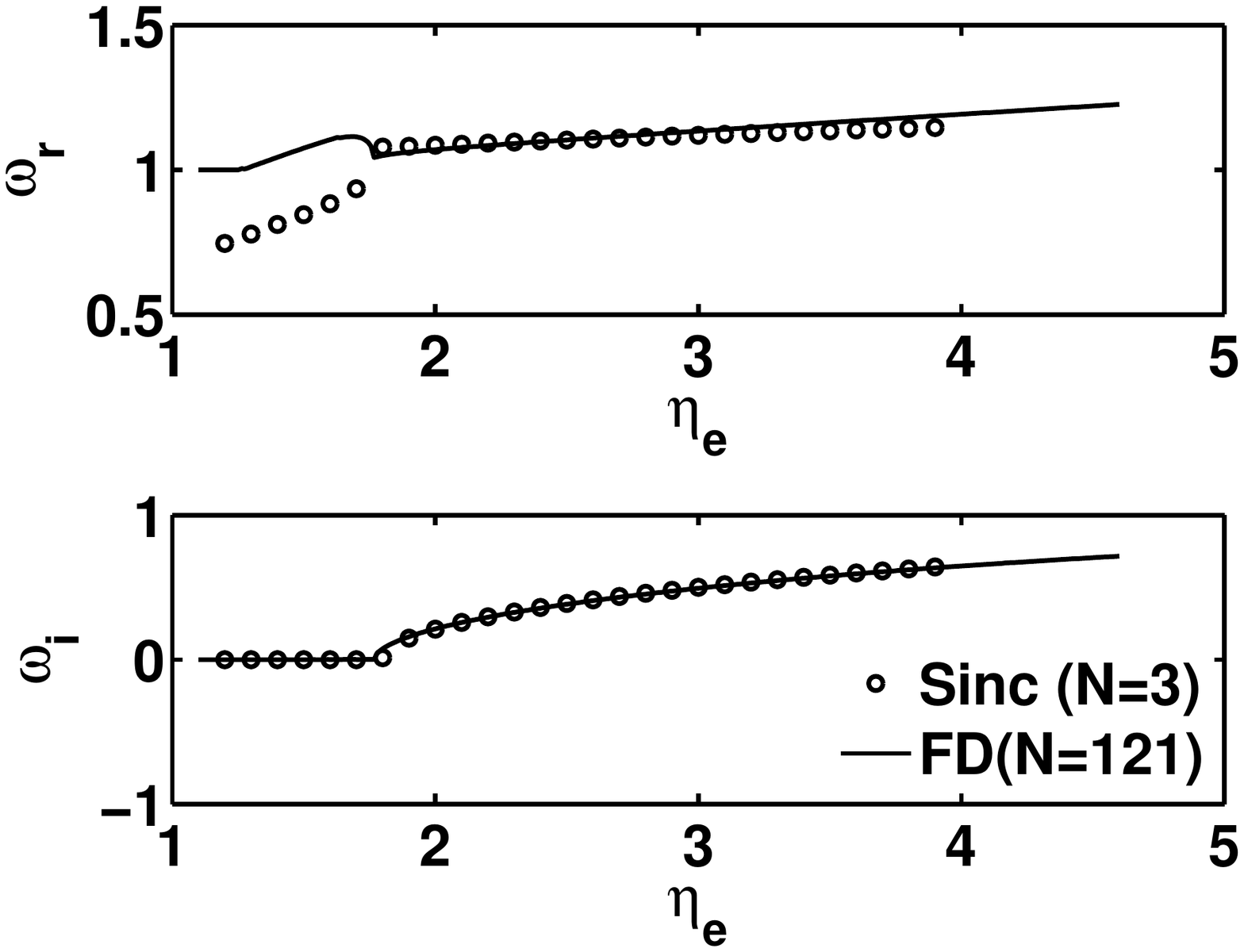}
\end{center} 
\caption{\label{sinc}}
\end{figure}

The use of Sinc method to compute the eigenvalues of the ETG mode has also been investigated.
The Sinc Method is usually applicable on an interval $(-\infty, \infty)$ using equidistant nodes
with spacing $h$ and the nodes
\begin{equation}
 x_k=(k-\frac{N+1}{2}) h, \quad k=1,..., N
\end{equation}
and the weight function$\alpha(x)=1$ and the interpolant
\begin{equation}
\phi_j(x)=\frac{sin[\pi (x-x_j)/h] }{\pi (x-x_j)/h}.
\end{equation}
Like the Hermite method discussed earlier, the Sinc method also contains a free parameter $h$. 
The results are encouraging and represented in Figure \ref{sinc}. It gives a reasonably accurate growth rate
for even N=3. Thus the Sinc method is very useful if one is
interested only in the computation of growth rates. 
The real frequency for $N=3$ is wrong by about $20\%$ but rapidly improves as $N$ increases.

\subsection{ Computation of the ITG growth rate}
\begin{figure}[p]
\begin{center}
\includegraphics[page=1, bb= 86   188   546   592]{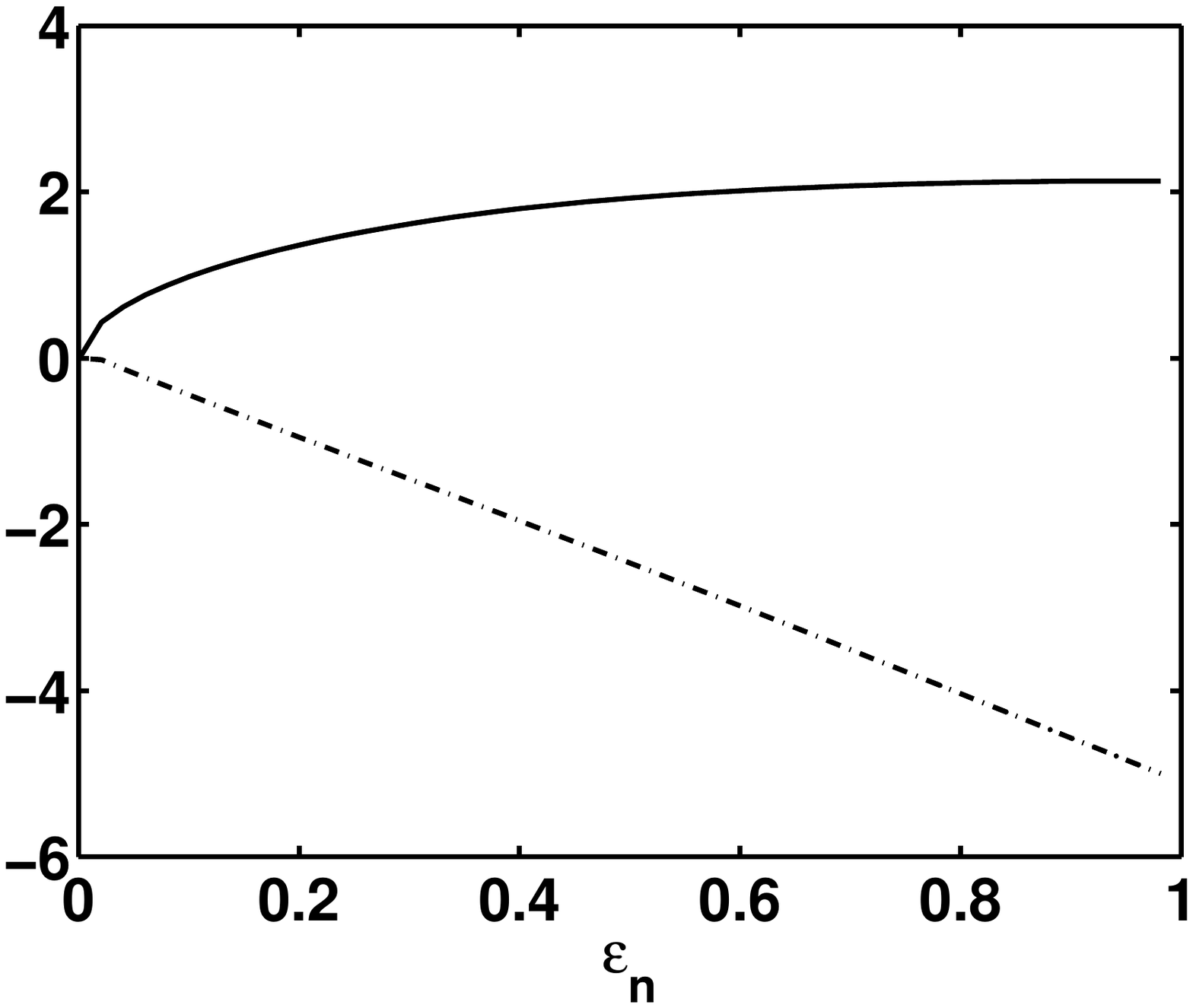}
\end{center} 
\caption{\label{grwthITG} }
\end{figure}

Apart from investigating the ETG mode in detail, some computations of the ITG
mode using a different set of basic equations has also been attempted.
The following are not a set of differentio-algebraic
equations as Eq. (\ref{FLeq}), but a single second order differential equation 
as derived in Ref. \cite{guo}:
\begin{equation}
\Omega \frac{d^2 \psi}{d \theta^2}=h \left[ (\Omega-1)A+2 \epsilon_n g(\theta) + b \Omega \right] \psi
\end{equation}
where
\begin{equation}
A=\frac{1+\frac{5}{3 \tau}\frac{2 \epsilon_n}{\Omega}g(\theta)}{F+\beta  \frac{2 \epsilon_n}{\Omega}g(\theta)}
\end{equation}
Where $\psi$ is the eigenvector proportional to the normalized eigenvector $e \phi/T_e$ as in Ref \cite{guo} .
The parameter $\theta$ is the extended ballooning coordinate, $q$ is the safety factor and $s=(r/q) dq/dr$ is 
the shear parameter, $g(\theta)=cos(\theta) + s \theta sin(\theta)$ is a function which will be symbolized
by $g$ from now on. The above equations can be converted into a polynomial eigenvalue problem 
of the form $\sum_{i=0}^{N} A_N \Omega^N X=0$ or
\begin{eqnarray}
C_3 \frac{d^2 \psi}{d \theta^2} + [C_5(-C_1+C_3C_4)\psi +C_2  \frac{d^2 \psi}{d \theta^2} ] \Omega \\ \nonumber
+ [C_5(-1+C_1 +C_2C_4 +C_3b) \psi ] \Omega^2
+\left[ C_5(1+b C_2) \psi \right] \Omega^3=0
\end{eqnarray}
where $C_1, C_2... C_5$ are independent of $\Omega$ and are defined as follows:
\begin{eqnarray}
C_1=\frac{10 \epsilon_n g}{3 \tau} \;  & \;   C_2=1+\frac{5}{3 \tau} \\ \nonumber
C_3=\frac{(1+\eta_i)}{\tau}-\frac{5}{3 \tau} + \beta 2 \epsilon_n g \; & \; C_4=2 \epsilon_n g \\  \nonumber
C_5=\frac{q^2 b_{\theta} \tau}{\epsilon_n^2} \nonumber 
\end{eqnarray}
The results of the numerical computation are depicted in Figure \ref{grwthITG} where the dashed line
corresponds to the growth rate and the dashed dotted line corresponds to the real frequency.
The parameters used are $\epsilon_n=1, \eta_i=8, q=1, \tau=1, b_{\theta}=0.1$ 
and $s=1$ and the boundary condition $\psi=0$ at $x= \pm \infty$.
As in the ETG case, the operator $d^2 \psi /d \theta^2$ is replaced by the corresponding Hermite 
differentiation matrix. 

\section{Summary and Remarks} \label{Concl}
Two methods to numerically solve for eigenvalues of the Electron Temperature Gradient (ETG)
mode were presented. These methods were Hermite and Sinc methods. For verification
purpose the growth rate was also computed analytically as well as numerically using finite difference methods.
The advantages and disadvantages of each of these methods were also highlighted.

Although these methods seem promising they also
suffer from some disadvantages. Firstly banded matrices of finite difference matrices are replaced by full matrices.
Secondly the scaling parameter $b$ in the Hermite method or the spacing $h$ in the Sinc method must be adjusted for optimum results.

For computing ETG eigenvalues, spectral collocation methods were found to be more accurate, faster and more
useful than finite difference methods. 
As another example, the eigenvalue equation for
the Ion Temperature Gradient (ITG) modes was also attempted. 
Furthermore, recenly, these methods were also applied to the  Drift Resistive Ballooning Modes (DRBM) with 
encouraging results\cite{TangriDRBM}.  
Finally, the results presented here are quite general 
and can be applied to a number of other microinstabilities.


{\bf Acknowledgements}

A major portion of this work had been done at Institute for Plasma Research, Bhat, Gandhinagar, India. The author 
is extremely thankful to Prof. Kaw and Prof. R. Singh for the guidance and useful 
discussions and suggestions and encouragement.

\newpage
\clearpage

\begin{center}
FIGURE CAPTIONS
\end{center}
\begin{enumerate}
\item[] FIG 1. Comparison of eigenvectors from different methods is shown. The analytical eigenvector from Eq. (\ref{AnalyticalVector}) is 
shown as a solid line and the numerical computation using the finite difference method 
as described in Sec \ref{sFD} is shown as dots (and labeled FD in legend). The solid squares refer to computation using the Hermite Method
as descibed in Sec. \ref{sDM}.

\item[] FIG 2. Real frequency $\omega_r$ and growth rate $\omega_i$ from finite difference method is compared with  
$\omega_r$ and $\omega_i$ from Hermite DM method. Note that Hermite Method with $N=16$ compares well with the finite 
difference method at $N=121$ when $\omega_i>0$ .

\item[] FIG 3. Real frequency $\omega_r$ and growth rate $\omega_i$ from the optimsed finite difference method (indicated by crosses)
is comapred with the finite difference method (triangles) and the Hermite DM which is indicated by solid dots.

\item[] FIG 4. Numerical convergence of ETG growth rate as a function of N. Here, the error 
$\delta(N) \equiv [{\omega_i(N) - \omega_i^h (N=100)}]/{\omega_i^h (N=100)}$ is plotted versus N
for three diffrent methods finite difference (triangles), the optimsed finite difference method (crosses) and the Hermite DM method (dots).

\item[] FIG 5. Numerical convergence of ETG growth rate is illustated through the 
congugacy of growth rates with the help of the ratio $\delta \gamma$ .

\item[] FIG 6. Ordinal and nearest distances plotted on a logarithmic 
scale versus mode number $j$ for the ETG mode, $\epsilon_n=0.9$.

\item[] FIG 7. Ordinal and nearest distances plotted on a logarithmic scale versus mode number j for the ETG mode, $\epsilon_n=0.5$.

\item[] FIG 8. Solution of the equations for the ETG real frequency $\omega_r$ and growth rate $\omega_i$ using Sinc Method is compared
with the finite difference method. Note the small value of $N=3$ compares well with $N=121$.

\item[]  FIG 9. Growth  rate (solid line) and real frequency (dashed-dotted line) of the ITG mode as a function of $\epsilon_n$.
\end{enumerate}

\newpage 
\begin{thebibliography}{999}
\bibitem{Jhowry} B.Jhowry and J.Anderson Phys.Plasmas, 10, 782 (2003).
\bibitem{Anderson} J. Anderson, H. Nordman, and J. Weiland Phys. Plasmas 8, 180 (2001). 
\bibitem{Shozo} Shozo Koshigoe and Arnold Tubis Phys. Fluids 30, 1715 (1987).
\bibitem{Yamagiwa} M. Yamagiwa, A. Hirose, and M. Elia Phys. Plasmas 4, 4031 (1997)
\bibitem{Raj} K. Rajendran, J. Q. Dong, and S. M. Mahajan Phys. Plasmas 4, 908 (1997)
 \bibitem{Boyd} J. P. Boyd {\it J. Comput. Phys.} {\bf 126} 11 (1996).
\bibitem{Canuto} C. Canuto, M. Y.  Hussaini, A. Quarteroni, T. A. Zang, 'pectral Methods in Fluid Dynamics' Springer-Verlag 1987
\bibitem{Tadmor} E. Tadmor, SIAM Journal on Numerical Analysis {\bf 23}, 1 (1986). 
\bibitem{SinghNF} R. Singh, P.K. Kaw and J. Weiland,   {\it Nucl. Fusion} {\bf 41}, 1219 (2001).
\bibitem{TangriETG} V. Tangri, R. Singh, and P. K. Kaw, Phys. Plasmas {\bf 12}, 072506 (2005).
\bibitem{mypop} R. Singh, V. Tangri, H. Nordman, and J. Weiland {\it et. al.} {\it Phys. Plasmas,} {\bf 8}, 4340 (2001).
\bibitem{HortonETG} W. Horton, B. G. Hong, and W. M. Tang, Phys. Fluids {\bf 31}, 2971 (1988)
\bibitem{gao} Z. Gao {\it et. al.}, {\it Phys. Plasmas,} {\bf 10}, 2831 (2003).
\bibitem{guo} S. C. Guo and J. Weiland {\it Nuclear Fusion,} {\bf 37}, 1095 (1997). 
\bibitem{dong} J. Q. Dong et al. {\it Phys. Plasmas,} {\bf 8}, 3635 (2001).  
\bibitem{Weid} J. A. C. Weideman and S. C. Reddy {\it ACM Transactions on Mathematical Software Archive} {\bf 26} 465 (2000).  
\bibitem{BoydBook} J. P. Boyd 'Chebyshev and Fourier Spectral Methods', Dover Publications, Second Edition Pg 1.
\bibitem{Schumer} J. W. Schumer {\it J. Comput. Phys.} {\bf 144} 626 (1998).
\bibitem{gary} J. Gary and R. Helgason  {\it J. Comput. Phys.} {\bf 5} 169 (1970).
\bibitem{Engel} F. Emgelmann {\it et. al.,} {\it Phys. Fluids} {\bf 6} 266 (1963). 
\bibitem{Antia} H. M. Antia, Book, Tata McGraw Hill Publishers
\bibitem{Numrecipes} W. H. Press , S. A. Teukolsky, W. T. Vetterling and B. P. Flannery {\it ``Numerical Recipes''}, Second Edition, Cambridge Press, (2000).
\bibitem{QZAlg} C. B. Moler and G. W. Stewart, SIAM J. Numer. Anal., {\bf 10} 241, (1973).
\bibitem{Jenko} F. Jenko, W. Dorland and G. W. Hammet {\it Phys. Plasmas,} {\bf 8} 4096 (2002). 
\bibitem{Lee} Y. C. Lee, J. Q. Dong, P. N. Guzdar and C. S. Liu   {\it Phys. Fluids,} {\bf 30}, 1331 (1987).
\bibitem{TangriDRBM} V. Tangri, A. Kritz, T. Rafiq and A. Pankin, APS $54^{th}$ Annual Meeting of the 
APS Division of Plasma Physics, October 29–November 2 2012; Providence, Rhode Island. 


\end{thebibliography}
\end{document}